\pdfoutput=1

\documentclass[11pt,twoside,a4paper,cmspaper,final,collab]{cms-tdr}
\def\svnVersion{218722}\def\cmsCernNo{2013-098}\def\cmsCernDate{\today}\def\cmsMessage{Published in the Journal of High Energy Physics as \href{http://dx.doi.org/10.1007/JHEP10(2013)164}{\doi{10.1007/JHEP10(2013)164.}}}

\begin{document}\cmsNoteHeader{SMP-12-020}

\hyphenation{had-ron-i-za-tion}
\hyphenation{cal-or-i-me-ter}
\hyphenation{de-vices}
\RCS$HeadURL: svn+ssh://svn.cern.ch/reps/tdr2/papers/SMP-12-020/trunk/SMP-12-020.tex $
\RCS$Id: SMP-12-020.tex 205760 2013-09-04 17:40:43Z alverson $
\newlength\cmsFigWidth
\ifthenelse{\boolean{cms@external}}{\setlength\cmsFigWidth{0.85\columnwidth}}{\setlength\cmsFigWidth{0.4\textwidth}}
\ifthenelse{\boolean{cms@external}}{\providecommand{\cmsLeft}{top}}{\providecommand{\cmsLeft}{left}}
\ifthenelse{\boolean{cms@external}}{\providecommand{\cmsRight}{bottom}}{\providecommand{\cmsRight}{right}}
\cmsNoteHeader{SMP-12-020}
\title{Measurement of the production cross section for
$\cPZ\gamma\rightarrow\cPgn\cPagn\gamma$
in pp collisions at $\sqrt{s}=7\TeV$ and limits on
$\cPZ\cPZ\gamma$ and $\cPZ\gamma\gamma$ triple gauge boson
couplings}

\date{\today}

\abstract{
A measurement of the
$\cPZ\gamma\to\cPgn\cPagn\gamma$ cross section in pp collisions at
$\sqrt{s} = 7\TeV$ is presented, using data corresponding to an integrated
luminosity of 5.0\fbinv collected with the CMS detector.
This measurement is based on the observation of events with
an imbalance of transverse energy in excess of 130\GeV and a single photon
in the absolute pseudorapidity range $\abs{\eta} < 1.4$ with
transverse energy above 145\GeV. The $\cPZ\gamma\to\cPgn\cPagn\gamma$
production cross section is measured to be
$21.1 \pm 4.2\stat \pm 4.3\syst \pm 0.5\lum\unit{fb}$,
which agrees with the standard model prediction of
$21.9 \pm 1.1\unit{fb}$. The results are combined with
the CMS measurement of Z$\gamma$ production in the $\ell^+\ell^-\gamma$
final state (where $\ell$ is an electron or a muon) to yield the most
stringent limits to date on triple gauge boson couplings:
$\abs{h_3^\cPZ}  < 2.7 \times 10^{-3}$,
$\abs{h_4^\cPZ}  < 1.3 \times 10^{-5}$ for $\cPZ\cPZ\gamma$ and
$\abs{h_3^\gamma}  < 2.9 \times 10^{-3}$, $\abs{h_4^\gamma}  < 1.5 \times 10^{-5}$
for $\cPZ\gamma\gamma$ couplings.
}

\hypersetup{%
pdfauthor={CMS Collaboration},%
pdftitle={Measurement of the production cross section for
Z gamma to nu nu gamma in pp collisions at sqrt(s) = 7 TeV and
limits on Z Z gamma and Z gamma gamma triple gauge boson couplings},
pdfsubject={CMS},%
pdfkeywords={CMS, physics}}

\maketitle

\section{Introduction}
The study of the production of pairs of gauge bosons in high-energy
particle collisions provides an important test
of the electroweak sector of the standard model~(SM). The diboson
production rate is
sensitive to the gauge boson self-interactions that arise as a consequence of
the non-Abelian nature of the SU(2)$\times$U(1) symmetry of the SM.
The values of these couplings are fixed in the SM, and any observed deviation
from the SM predictions would be an indication of new
physics at the vertex involving
the bosons~\cite{vvreview, baur_original}. In most searches for anomalous triple gauge couplings~(TGCs),
including this study, the
boson transverse energy spectrum is used as a sensitive observable,
as new physics is likely to result in an excess of energetic bosons
over the SM expectation.

In this paper, a study of the production of a Z boson and a photon, with
the Z boson decaying to a pair of neutrinos, is presented. We describe
the measurement of the production cross section as well as the extraction
of limits on the anomalous couplings for $\cPZ\cPZ\gamma$ and
$\cPZ\gamma\gamma$.

\begin{figure}[htb]
\begin{center}
\includegraphics[width=0.49\textwidth]{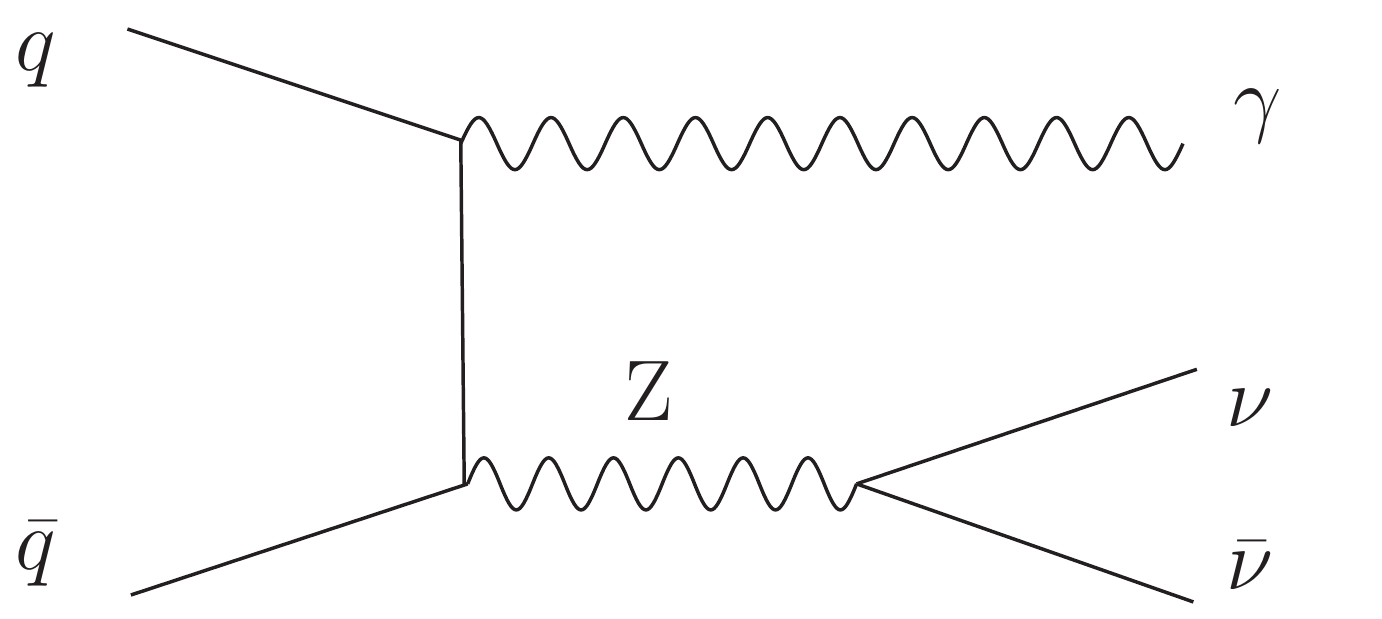}
\includegraphics[width=0.49\textwidth]{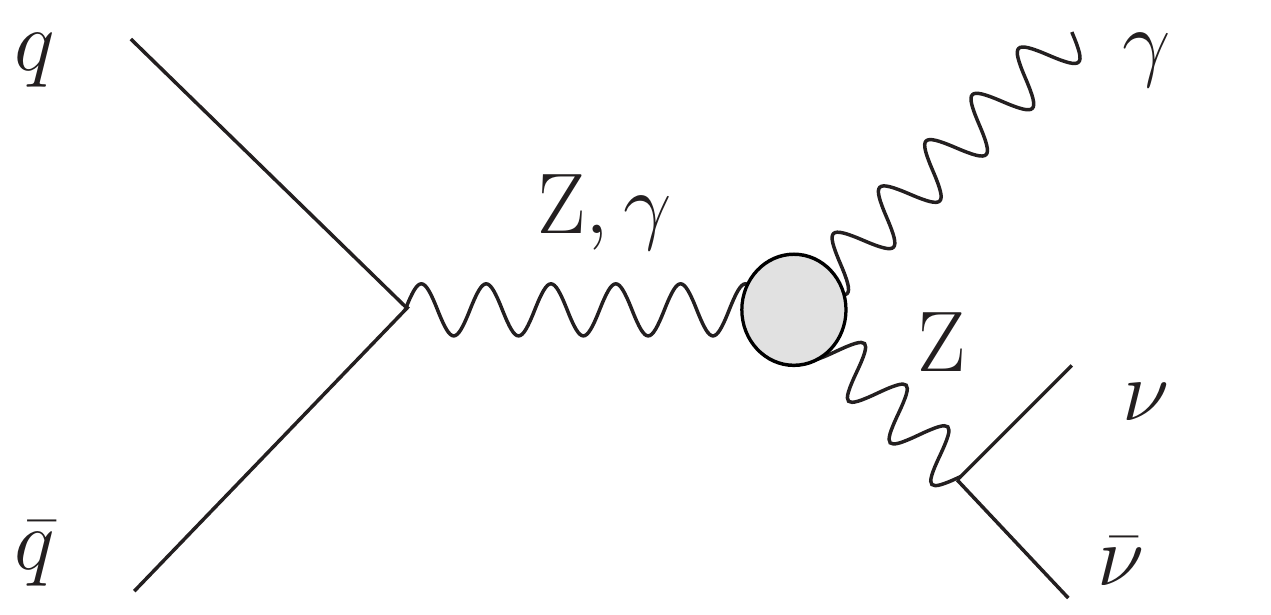}
\caption{Feynman diagrams of $\cPZ\gamma$ production via initial
state radiation in the SM at tree level (left) and via anomalous
$\cPZ\cPZ\gamma$ and $\cPZ\gamma\gamma$ triple gauge couplings
(right).}
\label{fig:feynman_zg}
\end{center}
\end{figure}

The $\cPgn\cPagn\gamma$ final state can be produced through initial-state
radiation, where a photon is emitted by an initial-state parton, or through
anomalous coupling vertices.
The allowed electroweak tree-level diagram in the SM for $\cPZ\gamma$
production in pp collisions is shown in Fig.~\ref{fig:feynman_zg}~(left).
The photon produced in the $s$ channel via an anomalous $\cPZ\cPZ\gamma$
or $\cPZ\gamma\gamma$ TGC is shown in Fig.~\ref{fig:feynman_zg}~(right).

The most general Lorentz- and gauge-invariant
$\cPZ\mathrm{V}\gamma$ vertex can be described by four coupling parameters
$h_{i}^\mathrm{V}$ $(i = 1, \ldots, 4)$~\cite{baur, baur1}, where V denotes
either a photon or a $\cPZ$ boson. The first two couplings ($i = 1, 2$) are
CP-violating, while the latter two are CP-conserving~\cite{baur1, baur1993}.
At tree level in the SM, the individual values of these TGCs are zero.
The photon transverse energy spectrum has similar sensitivity
to CP-violating and CP-conserving couplings~\cite{oldD0Result}.
Therefore, the results are generally interpreted in terms of CP-conserving
TGCs $h_3^\mathrm{V}$ and $h_4^\mathrm{V}$.

The sensitivity to TGCs in $\cPZ\gamma$ production
is higher in the $\cPZ\to\cPgn\cPagn$
decay mode than in Z boson decay modes with charged leptons for two reasons.
First, the branching fraction for a Z boson decay to a pair of neutrinos
is six times higher than for a decay to a particular charged lepton pair.
Second, the neutrino channel acceptance is higher.

Searches for anomalous $\cPZ\cPZ\gamma$ and
$\cPZ\gamma\gamma$ TGCs have been performed at
LEP~\cite{lepa,lepb,lepc}, the Tevatron~\cite{nunugammaFirstObservation,D0,cdf_pub},
and the Large Hadron Collider (LHC)~\cite{cms_result,atlas_result}.
No evidence has been found; the most stringent limits
in pp collisions at $\sqrt{s} = 7\TeV$ were set by
the ATLAS Collaboration~\cite{atlas_result}.

\section{CMS detector}
The measurements reported here are based on a data set corresponding
to an integrated luminosity of 5.0\fbinv, collected in 2011 at a
center-of-mass energy of $\sqrt{s}=7\TeV$
with the Compact Muon Solenoid (CMS) detector at the LHC.
The momenta of charged particles are measured by a silicon pixel
and strip tracker. The tracking system is surrounded by a crystal
electromagnetic calorimeter  and a brass-scintillator
hadron calorimeter, which measure particle energies deposited in
a barrel and two endcaps. The tracking system and calorimeters are
immersed in a 3.8\unit{T} magnetic field provided by the superconducting solenoid.
Muons are measured in gas-ionization detectors embedded in the steel
yoke outside the solenoid.
A more detailed description of the CMS detector can be
found in Ref.~\cite{CMS_det_jsnt}.
The trajectory of the particles in the detector can be described
by the azimuthal angle $\phi$, measured in the $x$-$y$ plane with respect to
the $x$ axis, and the pseudorapidity $\eta$, defined as
$\eta = -\ln\tan(\theta/2)$,
where $\theta$ is the polar angle of the trajectory of the particle
with respect to counterclockwise beam direction.

\section{Data selection and analysis}
Candidate events are selected in the experiment using unprescaled
single-photon triggers with the lower threshold on photon
transverse energy ranging from 75 to 135\GeV as defined by the two-level
CMS trigger system.
The triggers are fully efficient within 2\% statistical uncertainty
for photons with pseudorapidity
$\abs{\eta}  < 1.4$ and transverse energy $\ET^{\gamma} > 145\GeV$.

The final state consisting of an energetic photon accompanied
by an imbalance of transverse
energy can be mimicked by several other processes in the SM with at
least one photon in the final state, such as W$\gamma$, $\gamma$+jets, and
$\gamma\gamma$ processes, in which only a single energetic photon is
reconstructed in the event. Additionally, there are instrumental
backgrounds as discussed later in the paper.
Characteristics of both the signal and background events have been studied
using simulated events. A sample of signal events
$\cPZ\gamma\to \cPgn\cPagn\gamma$ is obtained using the
\PYTHIA v6.424~\cite{pythia} event generator at leading order.
The $\PW\gamma$ events are generated with the \MADGRAPH v5
generator~\cite{madgraph}, and the cross section is corrected to
include next-to-leading-order (NLO) effects
through a $K$ factor calculated with the \MCFM program~\cite{mcfm1, mcfm2}.
The $\gamma$+jets, W boson, and diphoton
samples are obtained using \PYTHIA. We do not consider final-state
radiation $\PW\to \ell\nu\gamma$ events in the \PYTHIA W boson sample
as those are included in the \MADGRAPH W$\gamma$ sample.
All leading-order simulation samples are propagated using CTEQ6L1~\cite{pdf_lhc}
parton distribution functions (PDFs).

To distinguish photons from misidentified jets, we apply
additional calorimetric selection criteria by requiring the shower shape to be
consistent with that of an electromagnetic particle~\cite{trigger_eff}.
To further reduce the contribution from processes with misidentified jets,
the photons are required to be isolated from other activity in
the electromagnetic and hadronic calorimeters as well as the tracker.
In particular, the scalar sum of the transverse component of the energy
deposits ($\ET$)  in the electromagnetic calorimeter within an
annulus of $0.06 <\Delta R< 0.40$ centered on the photon candidate, where
$\Delta R = \sqrt{(\Delta\eta)^2 + (\Delta\phi)^2}$,
must not exceed 5\GeV, while that in the hadronic calorimeter within
an annulus of $0.15 < \Delta R < 0.40$ must be below 2.6\GeV.
The scalar sum of the transverse momenta of charged particles, originating
from the primary vertex, in an annulus of $0.04 < \Delta R < 0.40$
must be below 2.1\GeV; the central cone around the photon,
$\Delta R < 0.04$, is excluded to maintain
a high efficiency for photons that initiate electromagnetic
showers early in the tracker.
Since the high instantaneous luminosity operation of the LHC results in
multiple pp interactions in a single bunch crossing (pileup),
the track isolation is calculated separately with respect to each vertex
in the event; the largest value of this isolation is then
required to satisfy the track isolation condition.
This procedure reduces the background from jets originating from
other vertices in the event corresponding to soft scattering processes.

Beam-halo muons are machine-induced particles that travel
parallel to the beam line. When such muons interact with the
electromagnetic calorimeter they can undergo a bremsstrahlung
process resulting in an electromagnetic energy
deposition identified as a photon.
To reduce the background due to bremsstrahlung photons from
beam-halo and cosmic-ray muons, the timing of the photon measured
in the CMS ECAL detector~\cite{timing} is required to be
consistent with that of the beam crossing. To further reduce
this background, additional requirements are imposed on the photon
candidates in events containing reconstructed muons.
Furthermore, to minimize the background due to the electrons from
$\PW\to \Pe\cPgn$ decays
that are misidentified as photons, the photons that have
a pattern of hits in the pixel detector similar to that expected
for electrons are rejected~\cite{pixel_seed}.

The imbalance of the transverse energy, $\MET$, in the reconstructed event
is defined by the
magnitude of the vector sum of the transverse energies of all the
reconstructed objects in the event, and is computed using a particle-flow
algorithm~\cite{pf_algo}.
The signal events have an $\MET$ comparable in
magnitude to the photon transverse energy because the neutrinos from
the Z boson decay recoil against the photon. In this study, events
are required to have $\MET > 130\GeV$, which is less stringent than the
145\GeV requirement on the photon $\ET$.
The $\MET$ criterion is fully efficient for signal events and
reduces a potential systematic uncertainty related to the modeling of
$\MET$ in the simulation.

Finally, events are vetoed if they contain other particles of significant
energy or momentum, defined by (i) a track with $\pt > 20\GeV$
that is $\Delta R > 0.04$ away from the photon or
(ii)~a jet, reconstructed with $\pt > 40\GeV$ using
the particles in the event identified by the particle-flow
algorithm and clustered using an anti-$\kt$
formalism~\cite{anti_kt_algo} with a
distance parameter of 0.5, within $\abs{\eta}  < 3.0$ and $\Delta R < 0.5$
of the axis of the photon.
After applying all of the selection criteria, 73 candidate events are selected.

The background from misidentified photons originating in jet
fragmentation and decay processes is estimated
by constructing a control sample in data, enriched with multijet events.
We use this sample to calculate a misidentification ratio, defined as
the number of events where the photon candidate satisfies the signal selection
criteria to the number of events where the photon candidate satisfies
looser selection criteria but fails the isolation condition.
Since not all photon candidates in the multijet control sample are from
misidentified jets, a correction is applied for
genuine photons from direct photon production in the multijet sample.
The photon contribution is estimated by
means of a fit to the shower shape profile of photon candidates in the
control data sample, where the distribution for genuine photons is
taken from simulated $\gamma$+jets events; this contribution is subtracted
from both the numerator and denominator of the misidentification ratio.
The corrected
ratio is then multiplied by the number of events in the signal data sample
that pass the photon selection criteria but fail the isolation requirements.
The background contribution due to misidentified jets is
estimated to be $11.2 \pm 1.6\stat \pm 2.2\syst$ events.
The systematic uncertainty reflects the modeling of the genuine photon
shower profile in the simulation, as well as the dependence of the
misidentification ratio on the choice of the looser selection criteria
and isolation requirements.

The dominant background where an electron is misidentified as a photon is
$\PW\to \Pe\nu$ production.
The contribution from this process is estimated from a control sample,
dominated by $\PW\to \Pe\nu$ events, that is selected by requiring events
to pass the full selection criteria except for the requirement of no
pixel hits for the photon. Then, the contribution to the $\cPZ\gamma$
sample from processes with misidentified electrons is estimated by scaling
the control sample by
$(1 - \epsilon_\text{ps})/\epsilon_\text{ps}$, where $\epsilon_\text{ps}$
is the probability for an electron to produce hits during its passage through
the pixel detector. The value of $\epsilon_\text{ps}$
is estimated from a $\cPZ\to\Pe\Pe$
data sample to be $\epsilon_\text{ps} = 0.994 \pm 0.003$.
The estimated number of events with misidentified electrons is
$3.5 \pm 0.1\stat \pm 1.5\syst$.
The systematic uncertainty is dominated by the uncertainty in measuring
$\epsilon_\text{ps}$ in $\cPZ\to\Pe\Pe$ events.

The total amount of background from
$\gamma$+jet, W$\gamma$, and diphoton processes,
after the full selection, is estimated from simulation to be
$4.4 \pm 1.0\stat \pm 0.4\syst$
events. The uncertainty in the simulated backgrounds takes
into account the uncertainty in the scale factor that corrects for
the data-simulation difference in the efficiency.

The out-of-time backgrounds, which are not produced via pp interactions, are
from beam-halo muons,
cosmic rays, and anomalous signals due to neutron interactions with
the electromagnetic calorimeter readout system
which can produce a
final-state signature of a single photon and an imbalance of transverse
energy. The contributions from these
sources are estimated with data from the timing and the shapes of the
shower energy deposition in the ECAL calorimeter
using the methods described in
Ref.~\cite{gamma_met_cms}. All contributions are found to be negligible
except for the beam-halo muon process, which is estimated to contribute
$11.1 \pm 0.6\stat \pm 5.5\syst$ events.
The uncertainty in the prediction is calculated by propagating
the uncertainty in the shape of the timing distribution for the beam-halo
bremsstrahlung shower.

\section{Cross section measurement}
The expected cross section of the signal process for
$\ET^\gamma > 145\GeV$ and $\abs{\eta} < 1.4$, obtained with
the NLO generator \textsc{wgrad}~\cite{baur1}, is $21.9 \pm 1.1\unit{fb}$.
A consistent prediction of $24.6 \pm 2.5 \unit{fb}$ is obtained
with \textsc{mcfm}. The quoted uncertainties in both
predictions take into account the PDF and scale uncertainties.
Based on the \textsc{wgrad} prediction, the expected number
of $\cPZ(\nu\cPagn)\gamma$ signal events
is $45.3 \pm 0.3\stat \pm 6.9\syst$.
The total number of expected background events is
$30.2 \pm  2.0\stat \pm 6.2\syst$.
A summary of the backgrounds and data yield is given in
Table~\ref{tab:BkgSummaryC}, wherein the uncertainties
in the background estimates include
both statistical and systematic sources.

\begin{table}[htbp]
\centering
\caption{Summary of estimated backgrounds to the $\cPZ(\nu\cPagn)\gamma$
process. The statistical and systematic uncertainties are added in quadrature.
\label{tab:BkgSummaryC}}
\begin{tabular}{|c|c|c|}
\hline
Source               & Number of selected events \\
\hline
Misidentified jets       & $11.2 \pm 2.8$ \\
Beam-halo muon processes & $11.1 \pm 5.6$ \\
Misidentified electrons  & $3.5  \pm 1.5$ \\
W$\gamma$                & $3.3  \pm 1.0$\\
$\gamma\gamma$           & $0.6 \pm 0.3$ \\
$\gamma$+jet             & $0.5 \pm 0.2$ \\
\hline
Total background                   & $30.2 \pm 6.5$ \\
\hline
$\cPZ\gamma \to \cPgn\cPagn\gamma$ (NLO) & $45.3 \pm 6.9$ \\
\hline
Data     & 73 \\
\hline
\end{tabular}
\end{table}

The $\cPZ\gamma \rightarrow \cPgn\cPagn\gamma$ cross section for
$\ET^{\gamma} > 145\GeV$ and $\abs{\eta}  < 1.4$ is calculated
using the following formulae:

$$ \sigma \times \mathcal{B}= \frac{N_\text{data}-N_\text{bg}}{A\times\epsilon\times L},$$
$$ A\times\epsilon = ({A\times\epsilon})_\text{sim} \times \rho,$$
where $N_\text{data}$ is the number of observed events, $N_\text{bg}$
is the estimated number of background events, $A$ is the geometrical and
kinematic acceptance of the selection criteria, $\epsilon$ is the
signal selection efficiency, and $L$ is the integrated
luminosity. The product of ${A\times\epsilon}$ is
estimated from the simulation to be $0.452 \pm 0.003$, where the
uncertainty is statistical. The
correction factor $\rho = 0.90 \pm 0.11$ takes into account
the efficiency difference between data and simulation for the trigger,
photon reconstruction and identification, consistency of cluster timing,
and jet and track veto requirements~\cite{gamma_met_cms}.

The largest contribution to the systematic uncertainty in
$(A\times\epsilon)_\text{sim}$ comes from the modeling of the photon
energy scale~\cite{trigger_eff}, which gives 4.3\%.
The uncertainties in the PDFs~\cite{pdf_lhc, PDF4LHC, MSTW08LO}
and the pileup each contribute 2.4\%. Additionally, there are
systematic uncertainties due to the energy scale and resolution in
the measurement of $\MET$~\cite{JME-10-009}, the jet energy scale
and resolution~\cite{PJetEnCor2011V2}, and the selection of the photon
production vertex; each of these contribute less than 3\%.
All these uncertainties are summarized
in Table~\ref{tab:AccSys}. The systematic uncertainty in the measured
integrated luminosity is 2.2\%~\cite{lumiPAS}.

\begin{table}[htbp]
\centering
\caption{Systematic uncertainties in $(A\times\epsilon)_\text{sim}$.
\label{tab:AccSys}}
\begin{tabular}{|c|c|}
\hline
Source & Systematic uncertainty in $({A\times\epsilon})_\text{sim}$ (\%)\\
\hline
Photon energy scale             & $\pm$ 4.3   \\
$\MET$ scale                & +1.6 $-$3.1   \\
$\MET$ resolution           & $\pm$ 0.03  \\
Jet energy scale                & $\pm$ 0.8   \\
Jet energy resolution           & $\pm$ 0.2   \\
Identification of photon vertex & $\pm$ 0.3   \\
Pileup modeling                & $\pm$ 2.4   \\
PDF modeling                    & $\pm$ 2.4   \\
\hline
Total                           & +5.7 $-$6.3   \\
\hline
\end{tabular}
\end{table}

The measured production cross section for $\cPZ\gamma \to \cPgn\cPagn\gamma$ for
$\ET^\gamma > 145\GeV$ and $\abs{\eta} < 1.4$
is $21.1 \pm 4.2\stat \pm 4.3\syst \pm 0.5\lum\unit{fb}$, which is in
agreement with the theoretical cross section, predicted at NLO,
of $21.9 \pm 1.1\unit{fb}$.

\begin{figure}
\begin{center}
\includegraphics[width=0.49\textwidth]{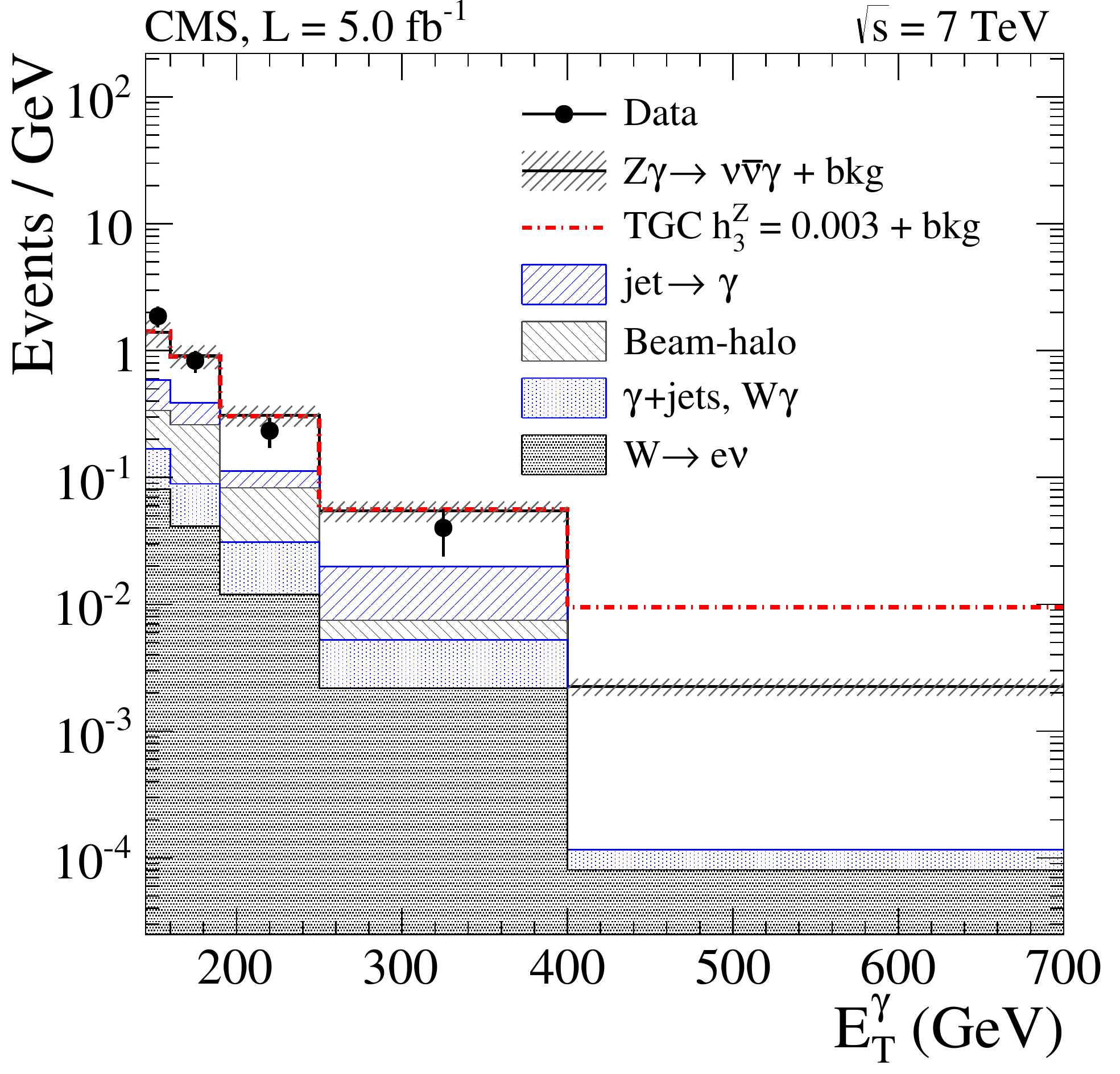}
\includegraphics[width=0.49\textwidth]{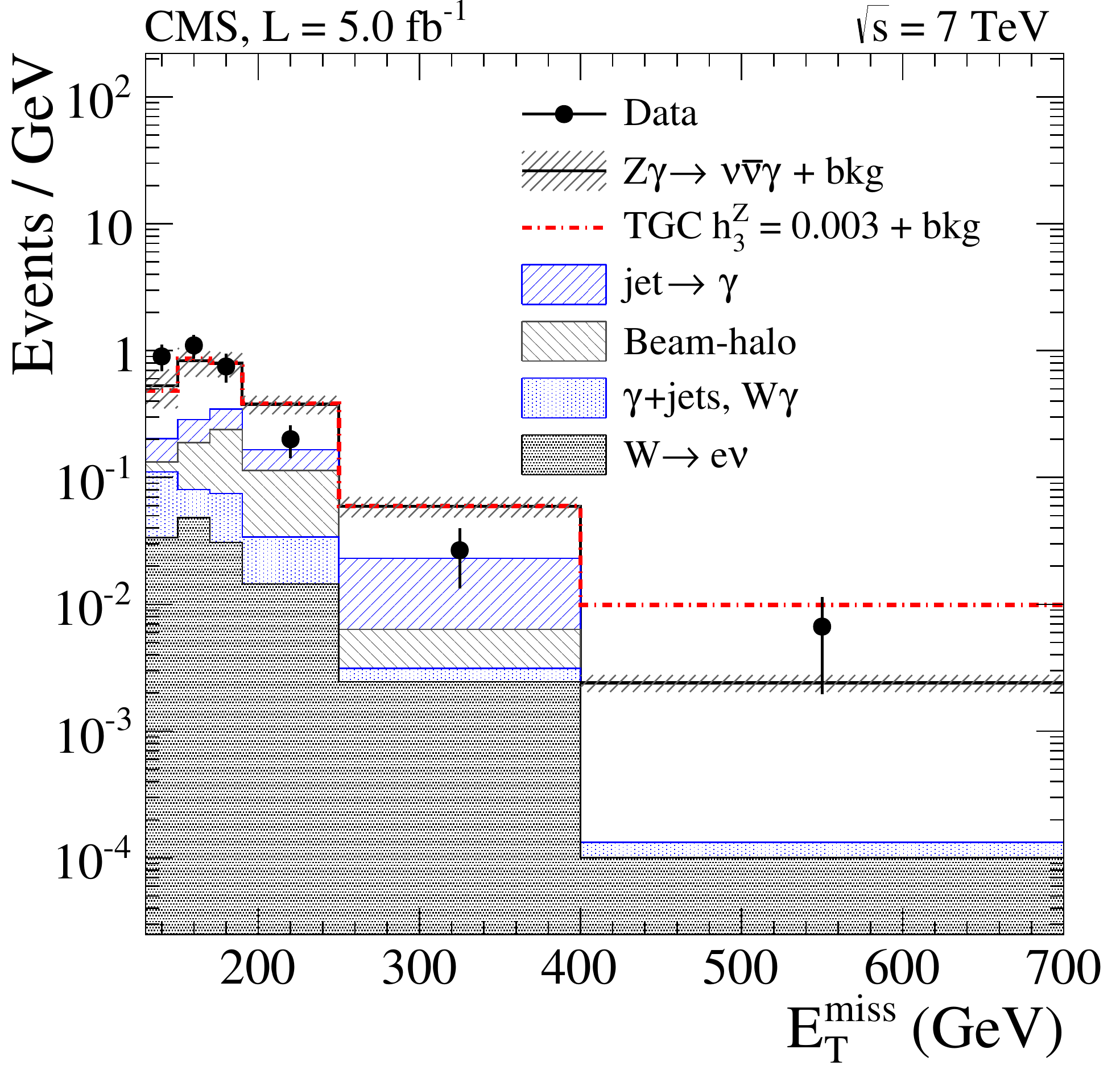}
\caption{The $\ET^\gamma$  and $\MET$ distributions in data
  (points with error bars) compared with the SM
  $\cPZ\gamma\to\cPgn\cPagn\gamma$ signal and
  estimated contributions from backgrounds (filled histograms). The shaded
  band illustrates the total uncertainty in the predicted distribution
  of the signal plus background hypothesis.
  A typical anomalous TGC signal would provide an excess, as shown in the
  dot-dashed histogram. The last bin in the distributions includes overflows.}
  \label{fig:photon_pt}
\end{center}
\end{figure}

The distributions of photon transverse energy and $\MET$ are given in
Fig.~\ref{fig:photon_pt}, with the signal and background predictions overlaid.
The expected contributions from a nonvanishing neutral TGC are also shown.
Such a coupling would give rise to more events with large $\ET^\gamma$
and $\MET$ than predicted in the SM. No excess of events is observed.

\section{Limits on triple gauge couplings}
We use the $\ET^\gamma$ spectrum to set limits on anomalous TGCs
by means of the likelihood formalism. Simulated samples
of $\cPZ\gamma$ signal for a grid of TGC values are produced
using the \SHERPA v1.2.2 generator~\cite{sherpa}.
In this study, we follow the CMS convention of not
suppressing the anomalous TGCs by an energy-dependent form factor.

The probability of observing the number of data events in a
given range of $\ET^\gamma$ is estimated using a
Poisson distribution given by the expected
signal and background predictions. The uncertainties in the quoted luminosity,
signal efficiency, and background fraction are considered to follow
a log-normal distribution. Limits on TGCs are calculated on the
basis of a modified frequentist CL$_\mathrm{s}$ method as described in
Refs.~\cite{CLs1, CLs2}.

\begin{figure}[htb]
\begin{center}
{\includegraphics[width=0.49\textwidth]{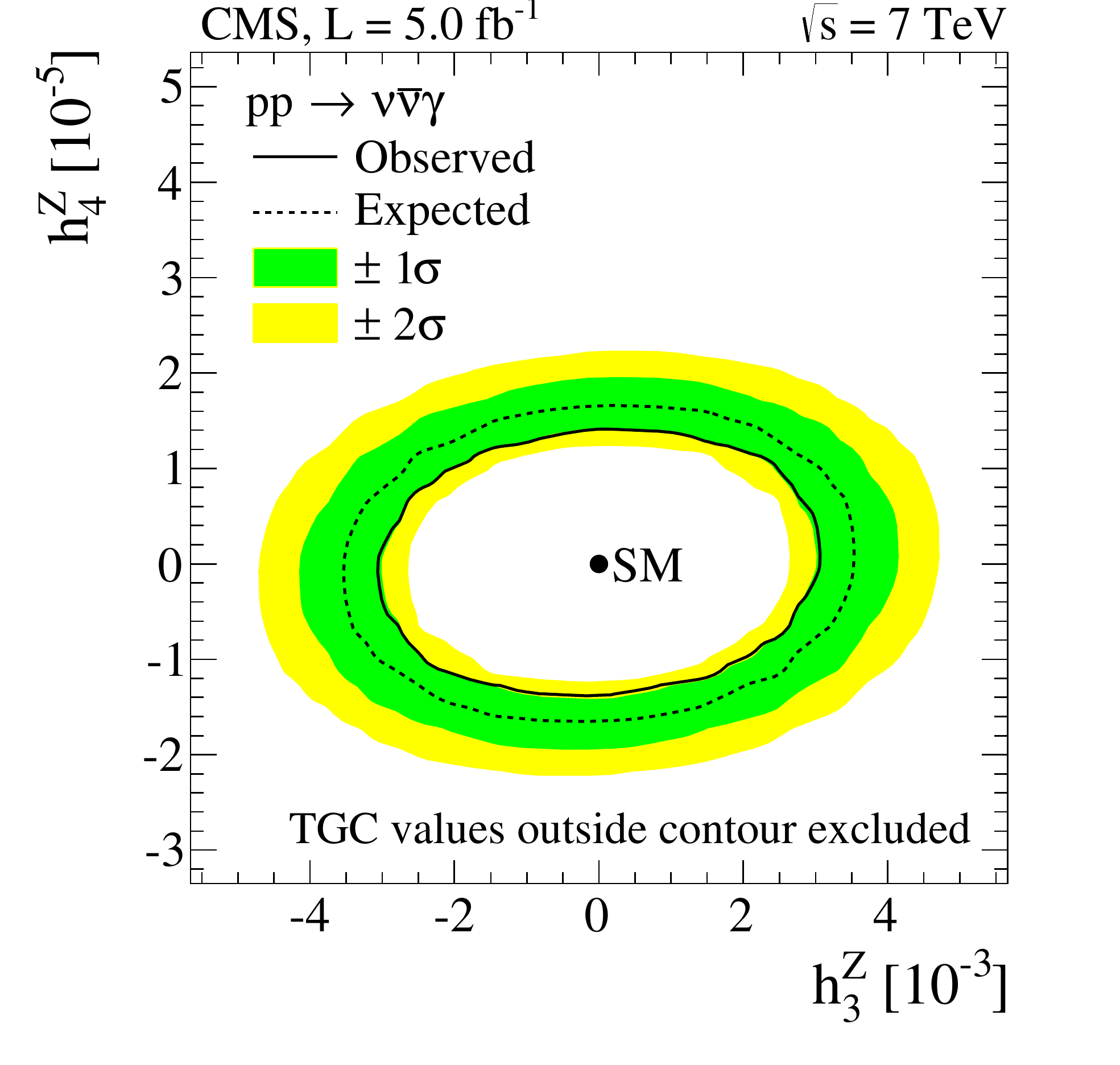}}
{\includegraphics[width=0.49\textwidth]{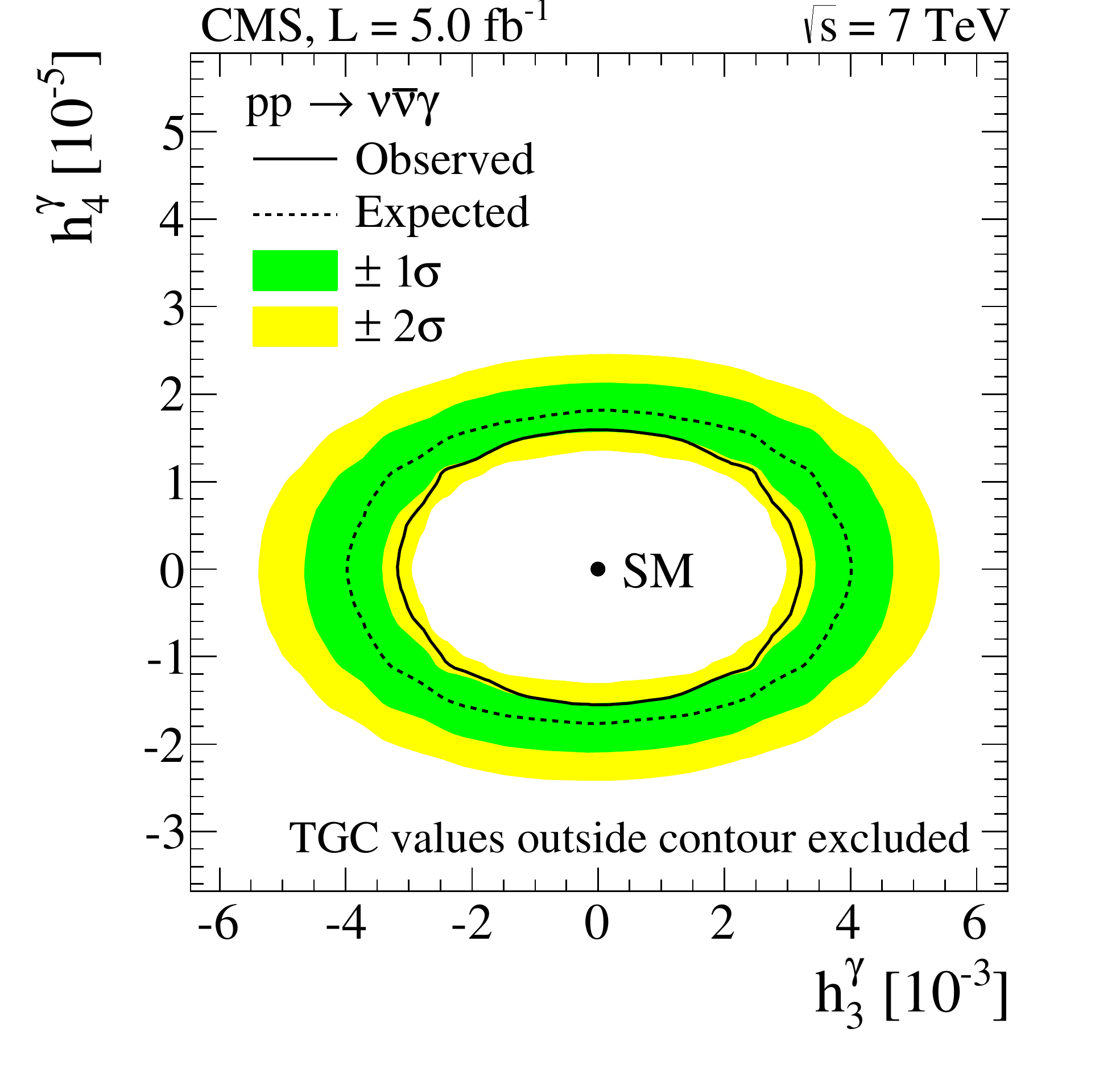}}
\caption{Two-dimensional 95\% CL limits on $\cPZ\cPZ\gamma$ couplings
  (left) and $\cPZ\gamma\gamma$ couplings~(right).}
\label{fig:2dlimits}
\end{center}
\end{figure}

\begin{figure}[htb]
\begin{center}
{\includegraphics[width=0.49\textwidth]{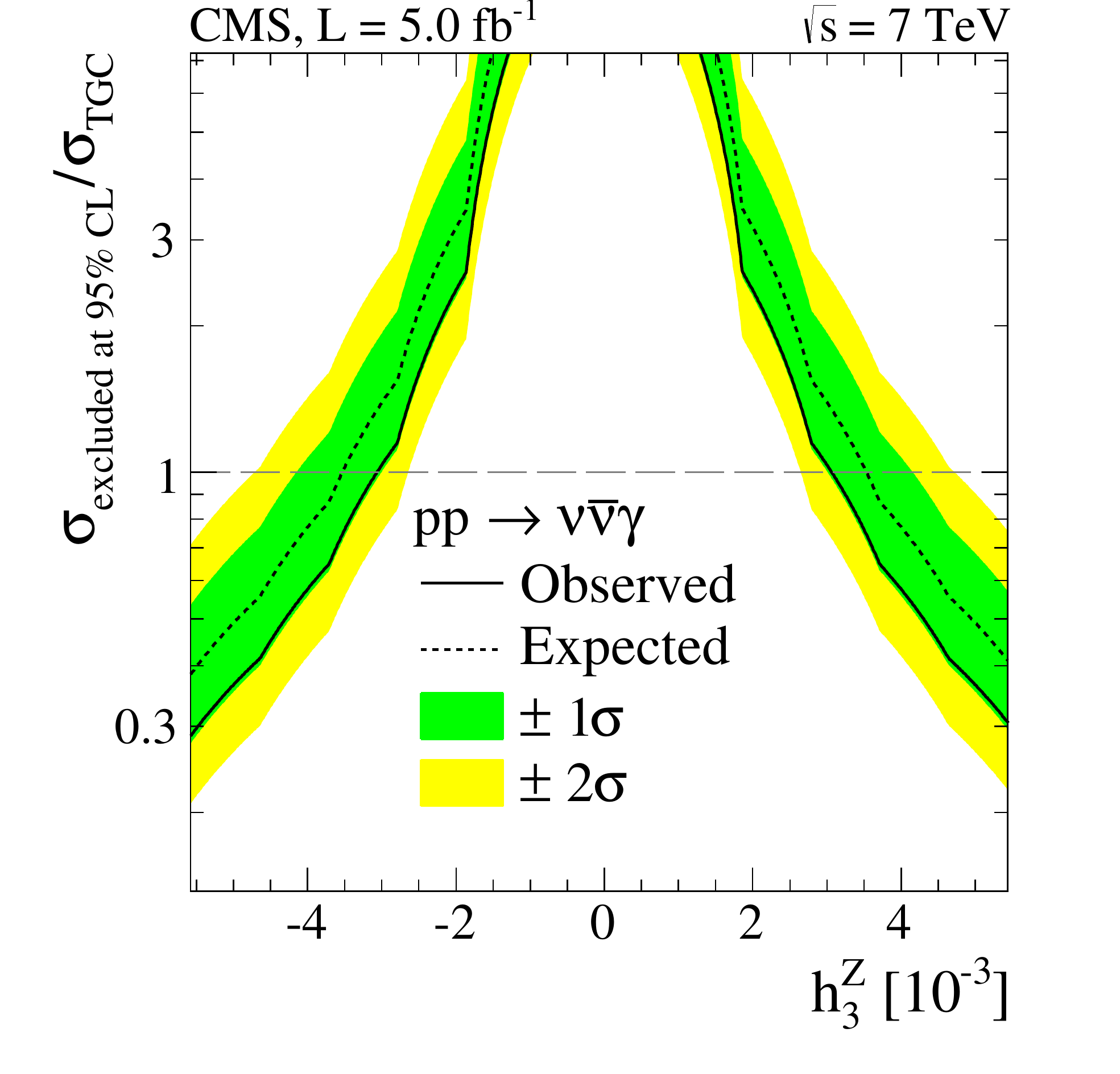}}
{\includegraphics[width=0.49\textwidth]{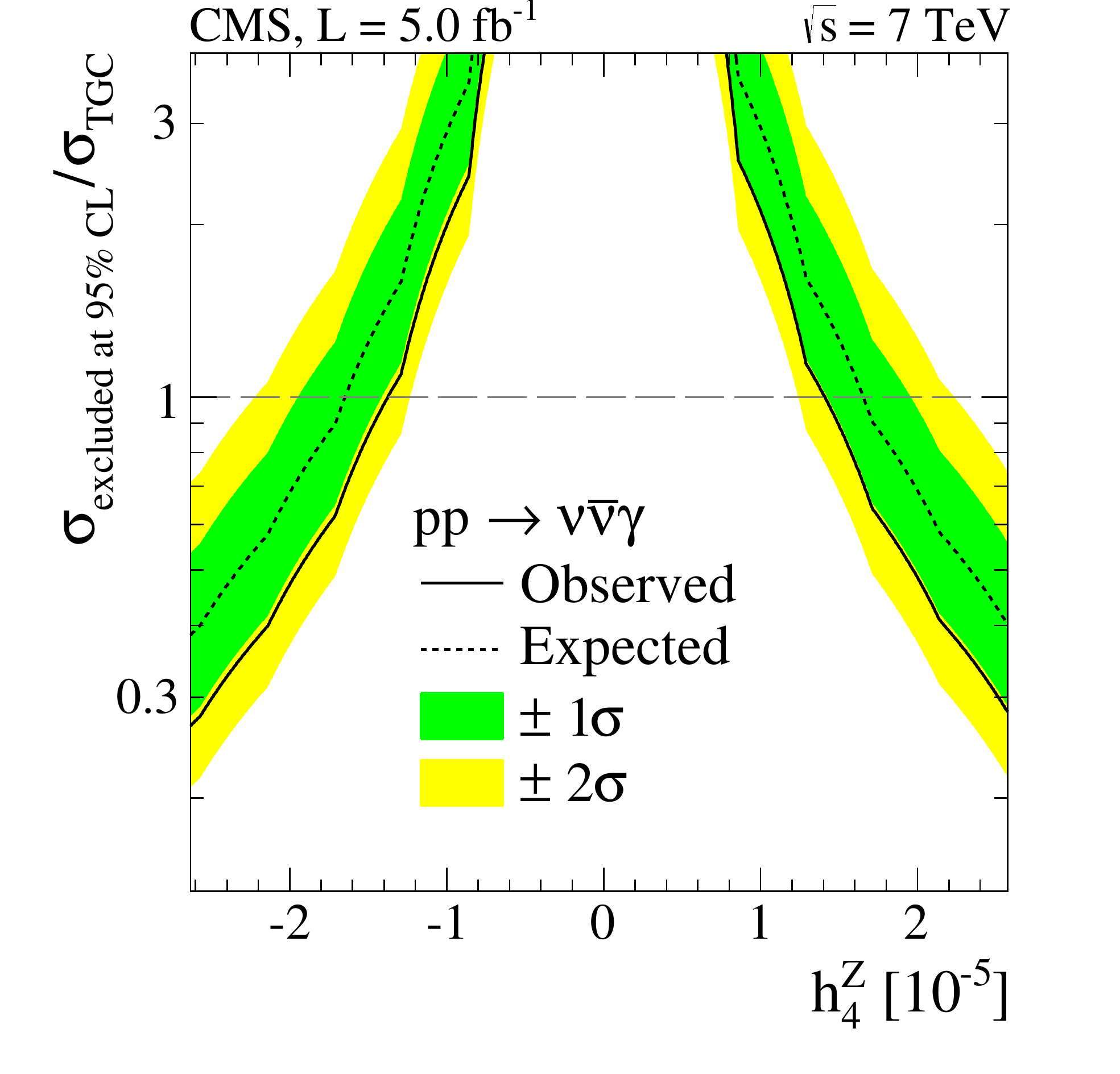}}
\caption{Cross section limits as functions of the
$\cPZ\cPZ\gamma$ couplings
$h_3^\cPZ$ (left) and $h_4^\cPZ$ (right).
The vertical axis represents the ratio of the 95\% CL upper limit
on the signal contribution from anomalous couplings to the expected
contribution for a given TGC hypothesis.}
\label{fig:limit_zzg_mono}
\end{center}
\end{figure}

\begin{figure}[htb]
\begin{center}
{\includegraphics[width=0.49\textwidth]{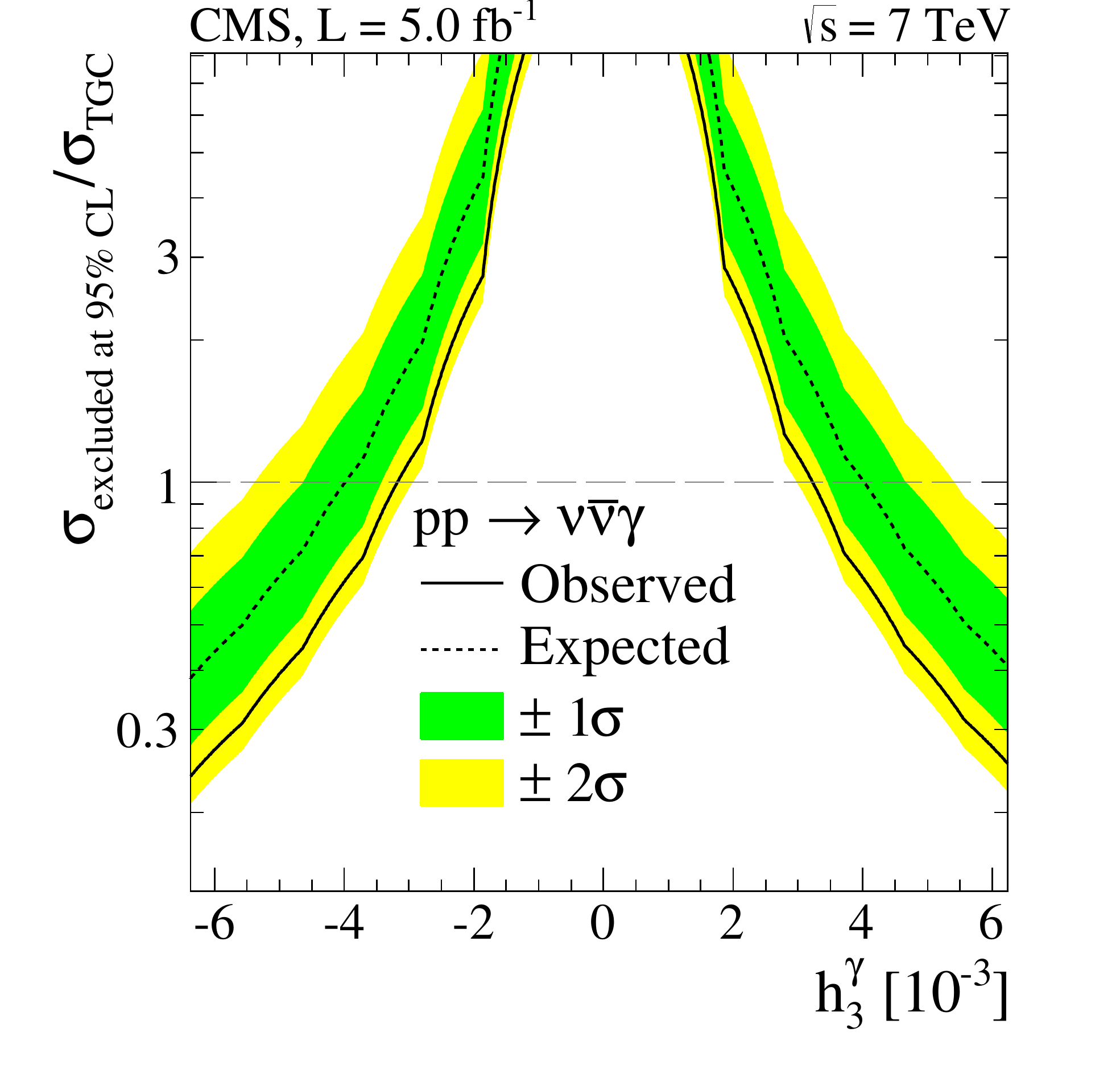}}
{\includegraphics[width=0.49\textwidth]{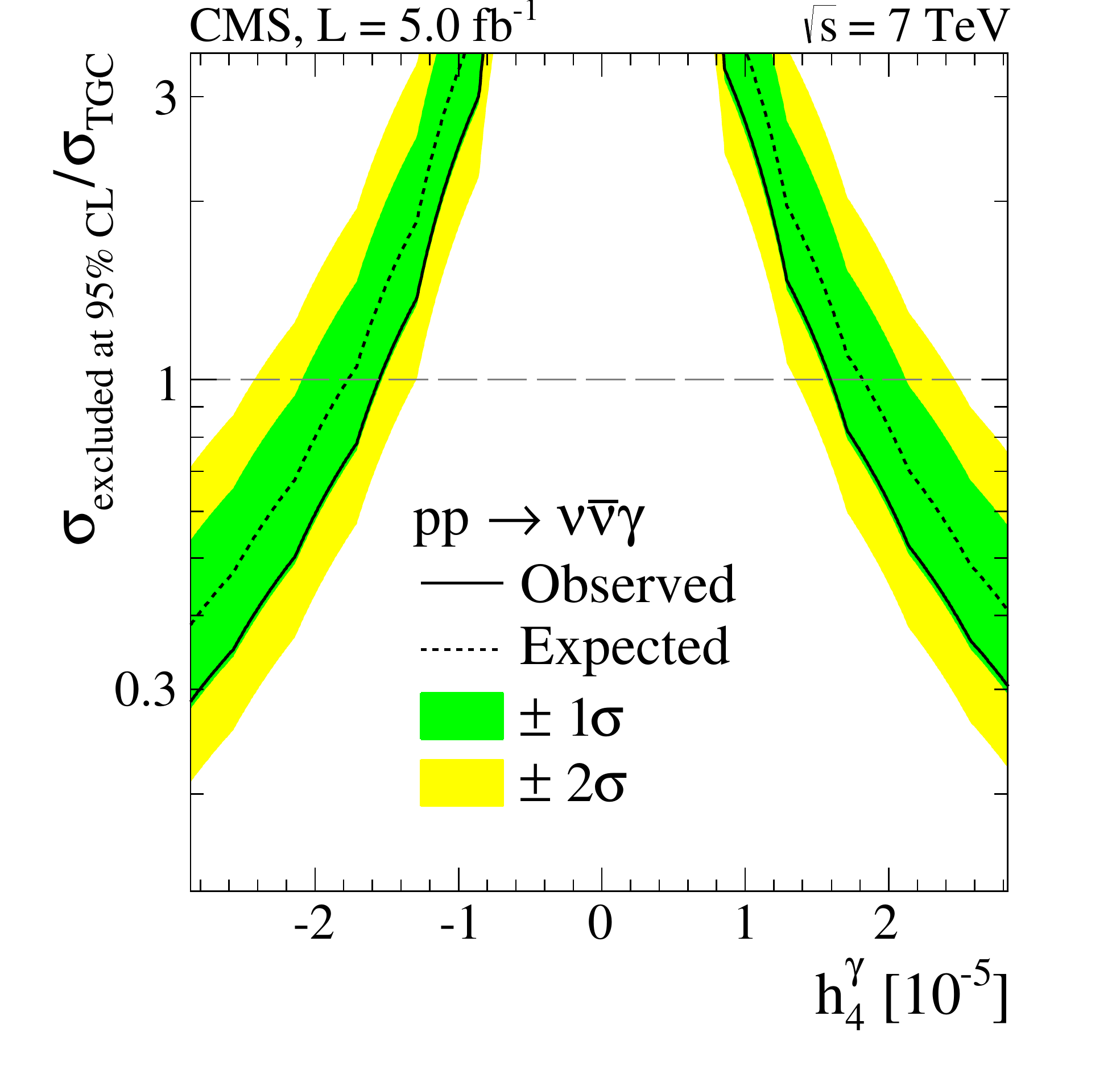}}
\caption{
Cross section limits as functions of the $\cPZ\gamma\gamma$ couplings
$h_3^\gamma$ (left) and $h_4^\gamma$ (right). The vertical axis represents
the ratio of the 95\% CL upper limit on the signal contribution from
anomalous couplings to the expected contribution for a given TGC
hypothesis.}
\label{fig:limit_zgg_mono}
\end{center}
\end{figure}

Limits at 95\% CL are set on pairs of TGC parameters $(h_{3}^{\cPZ}$,
$h_{4}^{\cPZ})$ and $(h_{3}^\gamma$, $h_{4}^\gamma)$, as presented in
Fig.~\ref{fig:2dlimits}.
Furthermore, one-dimensional 95$\%$ CL limits
are obtained for a given anomalous TGC while setting the other neutral TGCs
to their SM values, i.e., to zero.
The results, illustrated in
Figs.~\ref{fig:limit_zzg_mono}~and~\ref{fig:limit_zgg_mono}, are
$\abs{h_3^\cPZ}  < 3.1 \times 10^{-3}$,
$\abs{h_4^\cPZ}  < 1.4 \times 10^{-5}$,
$\abs{h_3^\gamma}  < 3.2 \times 10^{-3}$, and
$\abs{h_4^\gamma}  < 1.6 \times 10^{-5}$.

\begin{figure}[htb]
\begin{center}
{\includegraphics[width=0.49\textwidth]{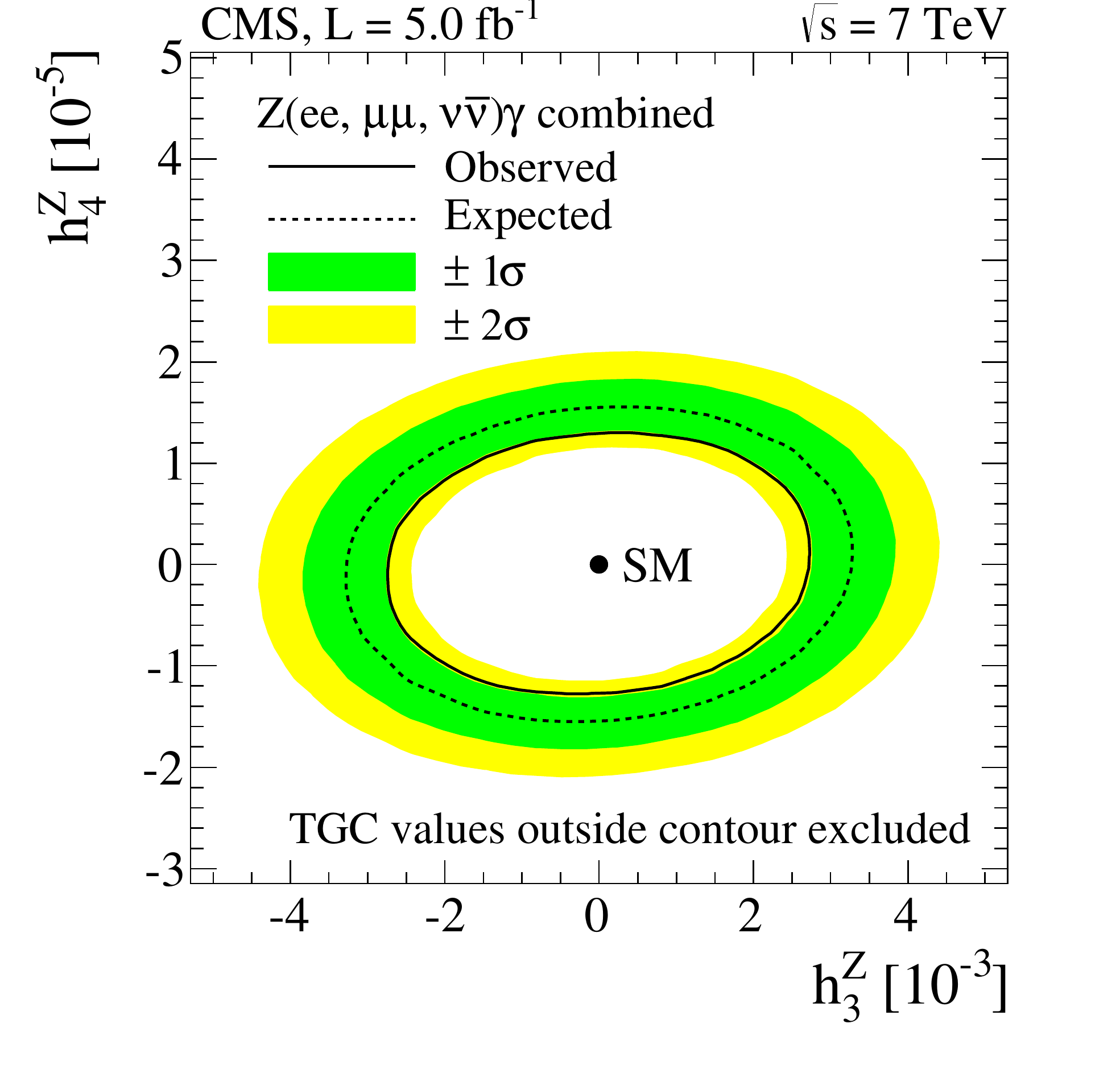}}
{\includegraphics[width=0.49\textwidth]{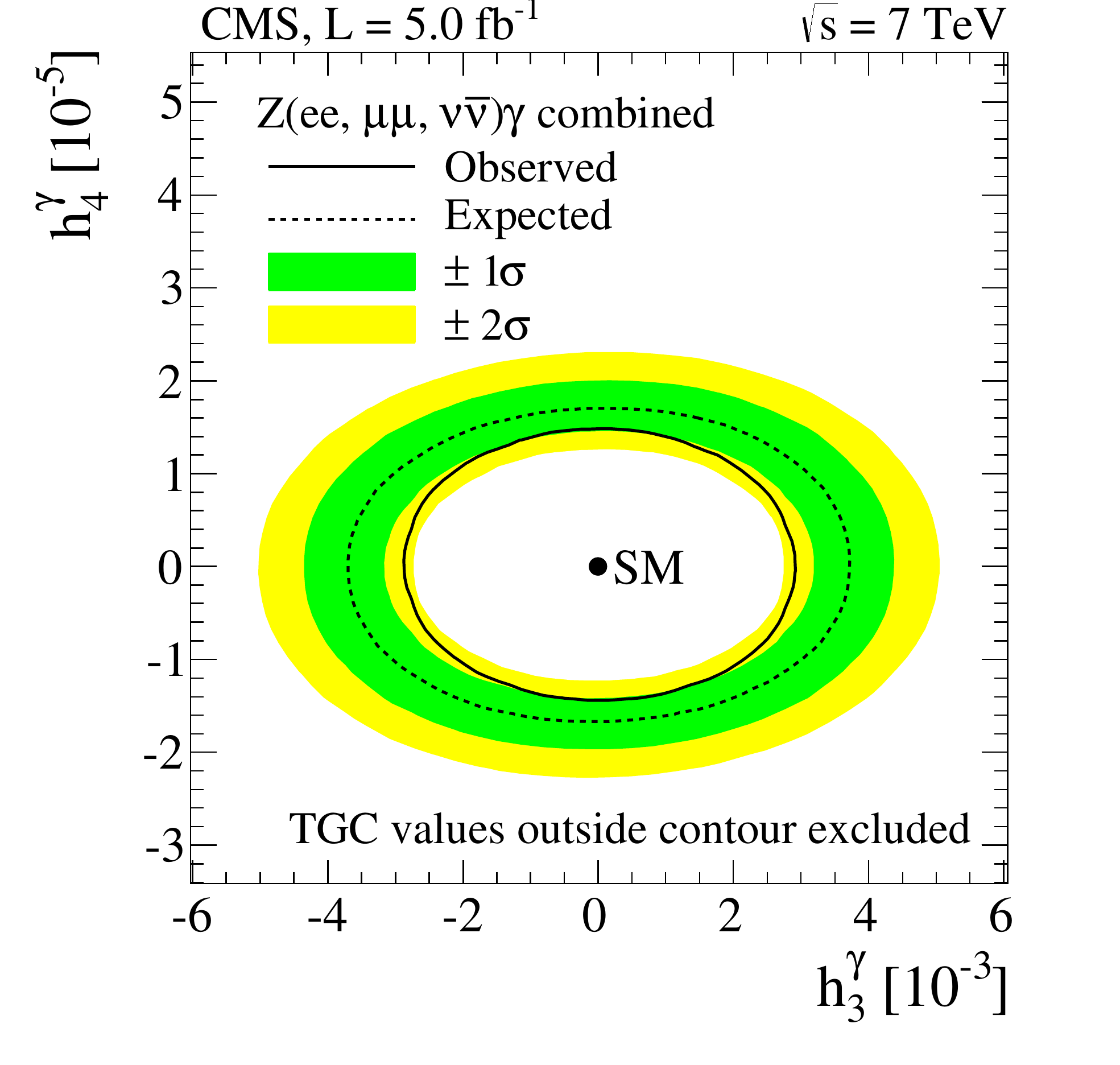}}
\caption{Two-dimensional 95\% CL limits on $\cPZ\cPZ\gamma$ couplings (left)
and $\cPZ\gamma\gamma$ couplings~(right) for combined neutral and charged
leptonic channels.}
\label{fig:2dlimitscombine}
\end{center}
\end{figure}

We combine the results presented here, based on the Z boson
decay to neutrinos, with those obtained by analysis of Z$\gamma$ candidate
events where the Z boson decays to a pair of electrons or
muons~\cite{wgammazgamma}. The combination is performed by summing the
negative log-likelihoods at each TGC hypothesis tested. The combined
limits are set in the same way as for each individual channel, while
also accounting for correlations between the systematic uncertainties
in different channels. The resulting two-dimensional 95\% CL bounds
on TGCs are given in Fig.~\ref{fig:2dlimitscombine}, and one-dimensional
limits are as follows:
$\abs{h_{3}^\cPZ } < 2.7 \times 10^{-3}$,
$\abs{h_{4}^\cPZ } < 1.3 \times 10^{-5}$,
$\abs{h_{3}^\gamma } < 2.9 \times 10^{-3}$, and
$\abs{h_{4}^\gamma} < 1.5 \times 10^{-5}$.

\section{Summary}
In conclusion, we have presented a measurement of the
$\cPZ\gamma\to \cPgn\cPagn\gamma$ production cross section in pp collisions
at $\sqrt{s} = 7\TeV$ in 5.0\fbinv of CMS data.
The measured $\cPZ(\nu\cPagn)\gamma$ cross section for
photons with $\ET>145\GeV$ and $\abs{\eta} <1.4$ is
$21.1 \pm 4.2\stat\pm 4.3\syst\pm 0.5\lum\unit{fb}$,
in agreement with the SM NLO prediction of
$21.9 \pm 1.1\unit{fb}$.
No evidence was found for anomalous neutral triple gauge couplings
in $Z\gamma$ production and 95\% CL limits have been placed on the
$h_3^\mathrm{V}$ and $h_4^\mathrm{V}$
parameters of $\cPZ\cPZ\gamma$ and $\cPZ\gamma\gamma$ couplings:
$\abs{h_3^\cPZ}  < 3.1 \times 10^{-3}$,
$\abs{h_4^\cPZ}  < 1.4 \times 10^{-5}$,
$\abs{h_3^\gamma}  < 3.2 \times 10^{-3}$, and
$\abs{h_4^\gamma}  < 1.6 \times 10^{-5}$.
These results, combined with those obtained from Z boson decays to
a pair of electrons or muons, yield the most stringent limits to date
on neutral triple gauge couplings:
$\abs{h_3^\cPZ}  < 2.7 \times 10^{-3}$,
$\abs{h_4^\cPZ}  < 1.3 \times 10^{-5}$ for $\cPZ\cPZ\gamma$ couplings and
$\abs{h_3^\gamma}  < 2.9 \times 10^{-3}$, $\abs{h_4^\gamma}  < 1.5 \times 10^{-5}$
for $\cPZ\gamma\gamma$. The results from the $\cPgn\cPagn\gamma$
analysis dominate the sensitivity to anomalous TGCs in Z$\gamma$ production.

\section*{Acknowledgements}
\hyphenation{Bundes-ministerium Forschungs-gemeinschaft Forschungs-zentren} We congratulate our colleagues in the CERN accelerator departments for the excellent performance of the LHC and thank the technical and administrative staffs at CERN and at other CMS institutes for their contributions to the success of the CMS effort. In addition, we gratefully acknowledge the computing centres and personnel of the Worldwide LHC Computing Grid for delivering so effectively the computing infrastructure essential to our analyses. Finally, we acknowledge the enduring support for the construction and operation of the LHC and the CMS detector provided by the following funding agencies: the Austrian Federal Ministry of Science and Research and the Austrian Science Fund; the Belgian Fonds de la Recherche Scientifique, and Fonds voor Wetenschappelijk Onderzoek; the Brazilian Funding Agencies (CNPq, CAPES, FAPERJ, and FAPESP); the Bulgarian Ministry of Education and Science; CERN; the Chinese Academy of Sciences, Ministry of Science and Technology, and National Natural Science Foundation of China; the Colombian Funding Agency (COLCIENCIAS); the Croatian Ministry of Science, Education and Sport; the Research Promotion Foundation, Cyprus; the Ministry of Education and Research, Recurrent financing contract SF0690030s09 and European Regional Development Fund, Estonia; the Academy of Finland, Finnish Ministry of Education and Culture, and Helsinki Institute of Physics; the Institut National de Physique Nucl\'eaire et de Physique des Particules~/~CNRS, and Commissariat \`a l'\'Energie Atomique et aux \'Energies Alternatives~/~CEA, France; the Bundesministerium f\"ur Bildung und Forschung, Deutsche Forschungsgemeinschaft, and Helmholtz-Gemeinschaft Deutscher Forschungszentren, Germany; the General Secretariat for Research and Technology, Greece; the National Scientific Research Foundation, and National Office for Research and Technology, Hungary; the Department of Atomic Energy and the Department of Science and Technology, India; the Institute for Studies in Theoretical Physics and Mathematics, Iran; the Science Foundation, Ireland; the Istituto Nazionale di Fisica Nucleare, Italy; the Korean Ministry of Education, Science and Technology and the World Class University program of NRF, Republic of Korea; the Lithuanian Academy of Sciences; the Mexican Funding Agencies (CINVESTAV, CONACYT, SEP, and UASLP-FAI); the Ministry of Business, Innovation and Employment, New Zealand; the Pakistan Atomic Energy Commission; the Ministry of Science and Higher Education and the National Science Centre, Poland; the Funda\c{c}\~ao para a Ci\^encia e a Tecnologia, Portugal; JINR, Dubna; the Ministry of Education and Science of the Russian Federation, the Federal Agency of Atomic Energy of the Russian Federation, Russian Academy of Sciences, and the Russian Foundation for Basic Research; the Ministry of Education, Science and Technological Development of Serbia; the Secretar\'{\i}a de Estado de Investigaci\'on, Desarrollo e Innovaci\'on and Programa Consolider-Ingenio 2010, Spain; the Swiss Funding Agencies (ETH Board, ETH Zurich, PSI, SNF, UniZH, Canton Zurich, and SER); the National Science Council, Taipei; the Thailand Center of Excellence in Physics, the Institute for the Promotion of Teaching Science and Technology of Thailand, Special Task Force for Activating Research and the National Science and Technology Development Agency of Thailand; the Scientific and Technical Research Council of Turkey, and Turkish Atomic Energy Authority; the Science and Technology Facilities Council, UK; the US Department of Energy, and the US National Science Foundation.

Individuals have received support from the Marie-Curie programme and the European Research Council and EPLANET (European Union); the Leventis Foundation; the A. P. Sloan Foundation; the Alexander von Humboldt Foundation; the Belgian Federal Science Policy Office; the Fonds pour la Formation \`a la Recherche dans l'Industrie et dans l'Agriculture (FRIA-Belgium); the Agentschap voor Innovatie door Wetenschap en Technologie (IWT-Belgium); the Ministry of Education, Youth and Sports (MEYS) of Czech Republic; the Council of Science and Industrial Research, India; the Compagnia di San Paolo (Torino); the HOMING PLUS programme of Foundation for Polish Science, cofinanced by EU, Regional Development Fund; and the Thalis and Aristeia programmes cofinanced by EU-ESF and the Greek NSRF.
\bibliography{auto_generated}   % will be created by the tdr script.

\providecommand{\href}[2]{#2}\begingroup\raggedright\begin{thebibliography}{10}%
\makeatletter
\providecommand{\hrefCMSnoop }[0]{\@secondoftwo}%
\makeatother
\providecommand{\doi}{\texttt{doi:}\begingroup \urlstyle{tt}\Url}

\bibitem{vvreview}
\hrefCMSnoop {} {J.~Ellison and J.~Wudka, ``Study of trilinear gauge-boson
  couplings at the Tevatron collider'',} \textit{ Ann. Rev. Nucl. Part. Sci.}
  \textbf{ 48} (1998) 33,
  \href{http://dx.doi.org/10.1146/annurev.nucl.48.1.33}{\doi{10.1146/annurev.nucl.48.1.33}}.

\bibitem{baur_original}
\hrefCMSnoop {} {K.~Hagiwara, R.~D. Peccei, and D.~Zeppenfeld, ``Probing the
  weak boson sector in $e^+ e^-\rightarrow W^+W^-$'',} \textit{ Nucl. Phys. B}
  \textbf{ 282} (1987) 253,
  \href{http://dx.doi.org/10.1016/0550-3213(87)90685-7}{\doi{10.1016/0550-3213(87)90685-7}}.

\bibitem{baur}
\hrefCMSnoop {} {G.~J. Gounaris, J.~Layssac, and F.~M. Renard, ``Signatures of
  the anomalous Z$\gamma$ and ZZ production at the lepton and hadron
  colliders'',} \textit{ Phys. Rev. D} \textbf{ 61} (2000) 073013,
  \href{http://dx.doi.org/10.1103/PhysRevD.61.073013}{\doi{10.1103/PhysRevD.61.073013}}.

\bibitem{baur1}
\hrefCMSnoop {} {U.~Baur and E.~L. Berger, ``Probing the weak boson sector in
  Z$\gamma$ production at hadron colliders'',} \textit{ Phys. Rev. D} \textbf{
  47} (1993) 4889,
  \href{http://dx.doi.org/10.1103/PhysRevD.47.4889}{\doi{10.1103/PhysRevD.47.4889}}.

\bibitem{baur1993}
\hrefCMSnoop {} {U.~Baur, S.~Errede, and J.~Ohnemus, ``Ratio of $W^{\pm}\gamma$
  and $Z\gamma$ cross sections: new tools in probing the weak boson sector at
  the Tevatron'',} \textit{ Phys. Rev. D} \textbf{ 48} (1993) 4103,
  \href{http://dx.doi.org/10.1103/PhysRevD.48.4103}{\doi{10.1103/PhysRevD.48.4103}}.

\bibitem{oldD0Result}
\hrefCMSnoop {} {{ D0} Collaboration, ``Study of ${\rm Z}\gamma$ events and
  limits on anomalous ${\rm ZZ}\gamma$ and ${\rm Z}\gamma\gamma$ couplings in
  $p\bar{p}$ collisions at $\sqrt{s} = 1.96\TeV$'',} \textit{ Phys. Rev. Lett.}
  \textbf{ 95} (2005) 051802,
  \href{http://dx.doi.org/10.1103/PhysRevLett.95.051802}{\doi{10.1103/PhysRevLett.95.051802}}.

\bibitem{lepa}
\hrefCMSnoop {} {{ L3} Collaboration, ``Study of $e^+e^{-}\rightarrow Z\gamma$
  process at LEP and limits on triple neutral-gauge-boson couplings'',}
  \textit{ Phys. Lett. B} \textbf{ 597} (2004) 119,
  \href{http://dx.doi.org/10.1016/j.physletb.2004.07.002}{\doi{10.1016/j.physletb.2004.07.002}}.

\bibitem{lepb}
\hrefCMSnoop {} {{ L3} Collaboration, ``Measurement of energetic single-photon
  production at LEP'',} \textit{ Phys. Lett. B} \textbf{ 346} (1995) 190,
  \href{http://dx.doi.org/10.1016/0370-2693(95)00023-E}{\doi{10.1016/0370-2693(95)00023-E}}.

\bibitem{lepc}
\hrefCMSnoop {} {{ OPAL} Collaboration, ``Study of Z pair production and
  anomalous couplings in ${\rm e^+}{\rm e^-}$ collisions at $\sqrt{s}$ between
  190 GeV and 209 GeV'',} \textit{ Eur. Phys. J. C} \textbf{ 32} (2003) 303,
  \href{http://dx.doi.org/10.1140/epjc/s2003-01467-x}{\doi{10.1140/epjc/s2003-01467-x}}.

\bibitem{nunugammaFirstObservation}
\hrefCMSnoop {} {{ D0} Collaboration, ``Measurement of the ${\rm
  Z}\gamma\to\nu\bar{\nu}\gamma$ cross section and limits on anomalous
  couplings in $p\bar{p}$ collisions at $\sqrt{s} = 1.96\TeV$'',} \textit{
  Phys. Rev. Lett.} \textbf{ 102} (2009) 201802,
  \href{http://dx.doi.org/10.1103/PhysRevLett.102.201802}{\doi{10.1103/PhysRevLett.102.201802}}.

\bibitem{D0}
\hrefCMSnoop {} {{ D0} Collaboration, ``$Z\gamma$ production and limits on
  anomalous $ZZ\gamma$ and $Z\gamma\gamma$ couplings in $p\bar p$ collisions at
  $\sqrt{s} = 1.96\TeV$'',} \textit{ Phys. Rev. D} \textbf{ 85} (2012) 052001,
  \href{http://dx.doi.org/10.1103/PhysRevD.85.052001}{\doi{10.1103/PhysRevD.85.052001}}.

\bibitem{cdf_pub}
\hrefCMSnoop {} {{ CDF} Collaboration, ``Limits on Anomalous Trilinear Gauge
  Couplings in Z$\gamma$ Events from $p\bar p$ Collisions at $\sqrt{s} =
  1.96\TeV$'',} \textit{ Phys. Rev. Lett.} \textbf{ 107} (2011) 051802,
  \href{http://dx.doi.org/10.1103/PhysRevLett.107.051802}{\doi{10.1103/PhysRevLett.107.051802}}.

\bibitem{cms_result}
\hrefCMSnoop {} {{ CMS} Collaboration, ``Measurement of W$\gamma$ and Z$\gamma$
  production in pp collisions at $\sqrt{s}= 7\TeV$'',} \textit{ Phys. Lett. B}
  \textbf{ 701} (2011) 535,
  \href{http://dx.doi.org/10.1016/j.physletb.2011.06.034}{\doi{10.1016/j.physletb.2011.06.034}}.

\bibitem{atlas_result}
\hrefCMSnoop {} {{ ATLAS} Collaboration, ``Measurement of ${\rm W}\gamma$ and
  ${\rm Z}\gamma$ production in pp collisions at $\sqrt{s}= 7\TeV$ with the
  ATLAS detector at the LHC'',} \textit{ Phys. Rev. D} \textbf{ 87} (2013)
  112003,
  \href{http://dx.doi.org/10.1103/PhysRevD.87.112003}{\doi{10.1103/PhysRevD.87.112003}}.

\bibitem{CMS_det_jsnt}
\hrefCMSnoop {} {{ CMS} Collaboration, ``The CMS experiment at the CERN LHC'',}
  \textit{ JINST} \textbf{ 3} (2008) S08004,
  \href{http://dx.doi.org/10.1088/1748-0221/3/08/S08004}{\doi{10.1088/1748-0221/3/08/S08004}}.

\bibitem{pythia}
\hrefCMSnoop {} {T.~Sj{\"{o}}strand, S.~Mrenna, and P.~Z. Skands, ``PYTHIA 6.4
  Physics and Manual'',} \textit{ JHEP} \textbf{ 5} (2006) 26,
  \href{http://dx.doi.org/10.1088/1126-6708/2006/05/026}{\doi{10.1088/1126-6708/2006/05/026}},
  \href{http://www.arXiv.org/abs/hep-ph/0603175}{\texttt{
  arXiv:hep-ph/0603175}}.

\bibitem{madgraph}
\hrefCMSnoop {} {J.~Alwall, M.~Herquet, F.~Maltoni, and O.~Mattelaer,
  ``MadGraph 5: Going Beyond'',} \textit{ JHEP} \textbf{ 6} (2011) 128,
  \href{http://dx.doi.org/10.1007/JHEP06(2011)128}{\doi{10.1007/JHEP06(2011)128}}.

\bibitem{mcfm1}
\hrefCMSnoop {} {J.~Campbell, R.~K. Ellis, and C.~Williams, ``Vector boson pair
  production at the LHC'',} \textit{ JHEP} \textbf{ 1107} (2011) 018,
  \href{http://dx.doi.org/10.1007/JHEP07(2011)018}{\doi{10.1007/JHEP07(2011)018}}.

\bibitem{mcfm2}
\hrefCMSnoop {} {J.~Campbell and R.~K. Ellis, ``An update on vector boson pair
  production at hadron colliders'',} \textit{ Phys. Rev. D} \textbf{ 60} (1999)
  113006,
  \href{http://dx.doi.org/10.1103/PhysRevD.60.113006}{\doi{10.1103/PhysRevD.60.113006}}.

\bibitem{pdf_lhc}
J.~Pumplin\hrefCMSnoop {} { {et~al.}, ``{New generation of parton distributions
  with uncertainties from global QCD analysis}'',} \textit{ JHEP} \textbf{ 07}
  (2002) 012,
  \href{http://dx.doi.org/10.1088/1126-6708/2002/07/012}{\doi{10.1088/1126-6708/2002/07/012}},
\href{http://www.arXiv.org/abs/hep-ph/0201195}{\texttt{ arXiv:hep-ph/0201195}}.
%%CITATION = HEP-PH/0201195;%%.

\bibitem{trigger_eff}
\href {http://cdsweb.cern.ch/record/1324545} {{ CMS} Collaboration, ``Isolated
  Photon Reconstruction and Identification at $\sqrt{s}=7$ {TeV}'',} CMS
  Physics Analysis Summary CMS-PAS-EGM-10-006, (2010).

\bibitem{timing}
\hrefCMSnoop {} {{ CMS} Collaboration, ``Time reconstruction and performance of
  the CMS electromagnetic calorimeter'',} \textit{ JINST} \textbf{ 5} (2010)
  T03011,
  \href{http://dx.doi.org/10.1088/1748-0221/5/03/T03011}{\doi{10.1088/1748-0221/5/03/T03011}},
  \href{http://www.arXiv.org/abs/0911.4044}{\texttt{ arXiv:0911.4044}}.

\bibitem{pixel_seed}
\href {http://cdsweb.cern.ch/record/1299116} {{ CMS} Collaboration, ``Electron
  Reconstruction and Identification at $\sqrt{s} = 7$ {TeV}'',} CMS Physics
  Analysis Summary CMS-PAS-EGM-10-004, (2010).

\bibitem{pf_algo}
\href {http://cdsweb.cern.ch/record/1279341} {{ CMS} Collaboration,
  ``Commissioning of the Particle-Flow Reconstruction in Minimum-Bias and Jet
  Events from {\Pp\Pp} Collisions at 7\TeV'',} CMS Physics Analysis Summary
  CMS-PAS-PFT-10-002, (2010).

\bibitem{anti_kt_algo}
\hrefCMSnoop {} {M.~Cacciari, G.~P. Salam, and G.~Soyez, ``The anti-$k_{T}$ jet
  clustering algorithm'',} \textit{ JHEP} \textbf{ 04} (2008) 063,
  \href{http://dx.doi.org/10.1088/1126-6708/2008/04/063}{\doi{10.1088/1126-6708/2008/04/063}}.

\bibitem{gamma_met_cms}
\hrefCMSnoop {} {{ CMS} Collaboration, ``Search for Dark Matter and Large Extra
  Dimensions in pp Collisions Yielding a Photon and Missing Transverse
  Energy'',} \textit{ Phys. Rev. Lett.} \textbf{ 108} (2012) 261803,
  \href{http://dx.doi.org/10.1103/PhysRevLett.108.261803}{\doi{10.1103/PhysRevLett.108.261803}}.

\bibitem{PDF4LHC}
M.~Botje\hrefCMSnoop {} { {et~al.}, ``{The PDF4LHC Working Group Interim
  Recommendations}'',} (2011).
  \href{http://www.arXiv.org/abs/1101.0538}{\texttt{ arXiv:1101.0538}}.

\bibitem{MSTW08LO}
\hrefCMSnoop {} {A.~D. Martin, W.~J. Stirling, R.~S. Thorne, and G.~Watt,
  ``Parton distributions for the LHC'',} \textit{ Eur. Phys. J. C} \textbf{ 63}
  (2009) 189,
  \href{http://dx.doi.org/10.1140/epjc/s10052-009-1072-5}{\doi{10.1140/epjc/s10052-009-1072-5}},
  \href{http://www.arXiv.org/abs/0901.0002}{\texttt{ arXiv:0901.0002}}.

\bibitem{JME-10-009}
\hrefCMSnoop {} {{ CMS} Collaboration, ``Missing transverse energy performance
  of the CMS detector'',} \textit{ JINST} \textbf{ 06} (2011) 9001,
  \href{http://dx.doi.org/10.1088/1748-0221/6/09/P09001}{\doi{10.1088/1748-0221/6/09/P09001}}.

\bibitem{PJetEnCor2011V2}
\href {http://cdsweb.cern.ch/record/1279362} {{ CMS} Collaboration, ``Jet
  Performance in pp Collisions at $\sqrt{s}$=7 {TeV}'',} CMS Physics Analysis
  Summary CMS-PAS-JME-10-003, (2010).

\bibitem{lumiPAS}
\href {http://cdsweb.cern.ch/record/1434360} {{ CMS} Collaboration, ``Absolute
  Calibration of the Luminosity Measurement at CMS : Winter 2012 Update'',} CMS
  Physics Analysis Summary CMS-PAS-SMP-12-008, (2012).

\bibitem{sherpa}
T.~Gleisberg\hrefCMSnoop {} { {et~al.}, ``Event generation with SHERPA 1.1'',}
  \textit{ JHEP} \textbf{ 02} (2009) 007,
  \href{http://dx.doi.org/10.1088/1126-6708/2009/02/007}{\doi{10.1088/1126-6708/2009/02/007}}.

\bibitem{CLs1}
\hrefCMSnoop {} {T.~Junk, ``Confidence level computation for combining searches
  with small statistics'',} \textit{ Nucl. Instrum. Meth. A} \textbf{ 434}
  (1999) 435,
  \href{http://dx.doi.org/10.1016/S0168-9002(99)00498-2}{\doi{10.1016/S0168-9002(99)00498-2}}.

\bibitem{CLs2}
\hrefCMSnoop {} {A.~L. Read, ``Presentation of search results: the CLs
  technique'',} \textit{ J. Phys. G} \textbf{ 28} (2002) 2693,
  \href{http://dx.doi.org/10.1088/0954-3899/28/10/313}{\doi{10.1088/0954-3899/28/10/313}}.

\bibitem{wgammazgamma}
\hrefCMSnoop {} {{ CMS} Collaboration, ``Measurement of the W$\gamma$ and
  Z$\gamma$ inclusive cross sections in pp collisions at $\sqrt{s} = 7\TeV$ and
  limits on anomalous triple gauge boson couplings'',} (2013).
  \href{http://www.arXiv.org/abs/1308.6832}{\texttt{ arXiv:1308.6832}}.
  Submitted to \textit{Phys. Rev. D}.

\end{thebibliography}\endgroup
\cleardoublepage \appendix\section{The CMS Collaboration \label{app:collab}}\begin{sloppypar}\hyphenpenalty=5000\widowpenalty=500\clubpenalty=5000\textbf{Yerevan Physics Institute,  Yerevan,  Armenia}\\*[0pt]
S.~Chatrchyan, V.~Khachatryan, A.M.~Sirunyan, A.~Tumasyan
\vskip\cmsinstskip
\textbf{Institut f\"{u}r Hochenergiephysik der OeAW,  Wien,  Austria}\\*[0pt]
W.~Adam, T.~Bergauer, M.~Dragicevic, J.~Er\"{o}, C.~Fabjan\cmsAuthorMark{1}, M.~Friedl, R.~Fr\"{u}hwirth\cmsAuthorMark{1}, V.M.~Ghete, N.~H\"{o}rmann, J.~Hrubec, M.~Jeitler\cmsAuthorMark{1}, W.~Kiesenhofer, V.~Kn\"{u}nz, M.~Krammer\cmsAuthorMark{1}, I.~Kr\"{a}tschmer, D.~Liko, I.~Mikulec, D.~Rabady\cmsAuthorMark{2}, B.~Rahbaran, C.~Rohringer, H.~Rohringer, R.~Sch\"{o}fbeck, J.~Strauss, A.~Taurok, W.~Treberer-Treberspurg, W.~Waltenberger, C.-E.~Wulz\cmsAuthorMark{1}
% ...
\vskip\cmsinstskip
% ...
1:~~Also at Vienna University of Technology, Vienna, Austria\\
2:~~Also at CERN, European Organization for Nuclear Research, Geneva, Switzerland\\
% ...

\end{sloppypar}
\end{document}